# Improving Urban Mobility by Understanding its Complexity
Carlos Gershenson

I have been working in the field of complex systems for almost two decades. As a computer engineer, I use simulations to explore scenarios for designing and controlling complex systems that can adapt to changes in their environment in a robust fashion. I have focussed on urban mobility because it deals with very complex systems and it affects billions of people worldwide. Moreover, with Mexico City having the "most painful commute" in the world, as a local I have an extra motivation to use recent scientific advances to improve mobility.

What makes mobility complex is that it is full of interactions. Interactions between pedestrians, between cars, between buses, between trains, between vehicles and infrastructure. In any mobility system, each component cannot be studied in isolation, as its future is partly but strongly determined by its interactions with other components and its environment. These interactions make it difficult to separate the components of a complex system. Traditional scientific and engineering methods rely on separability, and thus we need to use novel approaches. If we cannot study components individually, we need to model at the same time two levels of abstraction: the component level and the system level. We must understand how interactions between components give rise to system properties, and also how system properties constrain and promote behaviors and states of the components. Computer simulations have been the ideal tool for this, to the point that they have been compared with microscopes or telescopes which allow us to explore the microworld and the macroworld. Computer simulations allow us to explore the complex world.

As we have increased our understanding of complex systems, we have realized that interactions between components generate novel information that is not present in initial nor boundary conditions. This implies that even if we know everything about a complex system, their predictability is limited, as we do not know which information will be generated until it does. Science and engineering have assumed that the world is predictable, and that we just need to find the proper laws of nature to be able to foresee the future. But the study of complex systems has shown that this assumption is misguided. If novel information is produced by interactions, then the only way to reach the future is by actually going there. This limited prediction requires us to take a different approach when dealing with complex systems, such as those related to urban mobility. Instead of building predictive systems, we will be more efficient if we build adaptive systems that can adjust to the current situation at the same scale as it changes. There are things we can predict, and those we should, as it is convenient to deal with them beforehand. But we also know that there are things we cannot predict, and it will be better if we do not ignore them and provide our systems with the capabilities to adapt by themselves to the unexpected situations we can expect.

We can identify several factors that affect urban mobility: transportation requirements (living far from workplace or school), schedule distribution (if everyone has to be at the same place at the same time, the demand concentrates during rush hours), quantity (of passengers, vehicles), capacity (of public transport, infrastructure), technology (efficiency of infrastructure), planning and regulation (to avoid undesired situations, although they must be enforced), social contagion (if owning a private vehicle is seen as a sign of "success"), and human behavior (of passengers and drivers). The last one is perhaps the one citizens have most control on. Still, looking for individual benefits (trying to reach our destination faster), we can generate delays at the system level (e.g. blocking an intersection, or not letting doors close). Usually policies, regulations, and codes try to mediate between people precisely to avoid these conflicts. Still, in many cases people will seek the individual benefit as long as they can get away with it.

Certainly, we can identify some policies that are not efficient in all situations, tempting individuals to ignore the policies. For example, if on a freeway the speed limit is set so that it is safe to drive under all weather conditions, people will be tempted to break the limit if there are good conditions, especially if most drivers also do the same. Cameras and other sensors to detect and punish such devious behaviors work only locally, as drivers tend to change their speed only in their vicinity. A more effective approach would be to set dynamic speed limits according to the immediate situation. Moreover, due to the "slower-is-faster" effect, for dense traffic vehicles move faster if the speed limit is lower, as they can move continuously and avoid turbulent stop-and-go waves.

In other cases, policies are simply not understood by the public, making it difficult to achieve their adoption. If citizens are aware of the benefit their actions will have, they shall be more inclined to adopt a specific regulation. Surely, it has to be clear that the policy will bring benefits, which is not always the case. Unfortunately, many policies are the product of witticisms rather than scientific experimentation.

Based on my experience with complex systems and urban mobility, I can suggest the following five recommendations:

1. **Adaptation over prediction**. Urban systems change constantly. Even if we have all the positions and velocities of all the vehicles in a city, we cannot predict reliably for more than a couple of minutes into the future where vehicles will be, as their position will depend on so many externalities, such as the reaction times of other drivers, blocked lanes, pedestrians crossing, etc. We can have statistics about past densities and these can be useful for planning infrastructure, but our urban systems will be much more efficient if they can adapt to the changes in demand as fast as these occur, i.e. within seconds.
2. **Regulate interactions**. One way to achieve efficient mobility is by regulating interactions of the components of a system. If the behavior of one component affects negatively the mobility of another, we can say that there is friction generated. If we can regulate interactions to minimize friction, we will achieve efficient performance. This is also evident with the slower-is-faster effect: if components try to maximize their benefit, in many cases they create negative interactions which lead to global inefficiency. If we regulate and constrain the components, even if they do not go as fast as they would like to, they can reach their destination faster, so everyone benefits (components and system).
3. **Use sensors**. To make correct decisions, systems require information. Sensors are becoming cheaper, so we can massively deploy them to obtain relevant information. By relevant, I mean the necessary information to be able to adapt to changes in demand as they occur.
4. **Use algorithms**. Information collected by sensors can sit nicely on the cloud. But to make use of it, adaptive algorithms are required which are able to respond precisely to the changes in demand. In our laboratory, we have used self-organization to design adaptive algorithms: instead of trying to solve a problem which we know will change in ways we do not know, we build components that will constantly seek by their interactions solutions to the current situation. So when the situation changes, the algorithm adapts.
5. **Use agents**. If algorithms can give us solutions, these have to be taken into the real world. In some cases, agents are already there, e.g. traffic lights. But in others we still have to design them, e.g. to regulate driver or passenger behaviors. Agents must have the ability to influence the urban mobility systems towards the desired state, otherwise sensors and algorithms will be of little use.

I am optimistic about the future of urban mobility. We have many challenges ahead, but we can already see potential solutions that we can try and learn from. I do not want future generations to suffer by having mobility as a constraint. I believe we can make mobility to be rather an opportunity.